\documentclass[11pt]{article}
\usepackage{amsmath}
\usepackage{amsfonts}
\newcounter{mnotecount}[section]

\renewcommand{\themnotecount}{\thesection.\arabic{mnotecount}}

\newcommand{\mnote}[1]
{\protect{\stepcounter{mnotecount}}$^{\mbox{\footnotesize
$
\bullet$\themnotecount}}$ \marginpar{
\raggedright\tiny\em
$\!\!\!\!\!\!\,\bullet$\themnotecount: #1} }

\def\be{\begin{equation}}
\def\ee{\end{equation}}
\def\bea{\begin{eqnarray}}
\def\eea{\end{eqnarray}}

\newcommand{\HH}{{\bf H}}
\newcommand{\GG}{{\bf G}}
\newcommand{\uu}{{\bf u}}
\newcommand{\bbR}{\mathbb{R}}

\newcommand{\dive}{\operatorname{div}}
\newcommand{\curl}{\operatorname{curl}}
\newcommand{\grad}{\operatorname{grad}}

\newtheorem{Teo}{Theorem}[section]
\newtheorem{Prop}[Teo]{Proposition}

\newtheorem{Rem}[Teo]{Remark}

\def\eqq{\stackrel{\sigma}{=}}
\begin{document}
\title{Matching stationary spacetimes}
\author{Filipe C. Mena$^{1,2}$ and Jos\'e Nat\'ario$^{2}$\\
{\small $^1$ Departamento de Matem\'atica,
Universidade do Minho,
4710-057 Braga, Portugal}\\ 
{\small $^2$ Departamento de Matem\'atica, Instituto Superior T\'ecnico, 1049-001 Lisboa, Portugal}}
\date{19 February 2008}
\maketitle
\begin{abstract} 
Using the quasi-Maxwell formalism, we derive the necessary and sufficient conditions for the matching of two stationary spacetimes along a stationary timelike hypersurface, expressed in terms of the gravitational and gravitomagnetic fields and the $2$-dimensional matching surface on the space manifold. 
We prove existence and uniqueness results to the matching problem for stationary perfect fluid spacetimes with 
spherical, planar, hyperbolic and cylindrical symmetry. Finally, we find an explicit interior for the cylindrical analogue of the NUT spacetime. 
\end{abstract}   
\section*{Introduction}
There are hardly any general results on the existence and uniqueness of solutions to the matching problem in General Relativity. This is related to the fact that little is known about the initial boundary value problem for the Einstein equations \cite{Nagy-Fried}.

Existence and uniqueness of a vacuum exterior has been proved in certain symmetric cases, e.g.~for interiors with static spherical symmetry \cite{Kind-Ehlers}, homogeneous cylindrical symmetry \cite{Tod-Mena} and stationary axial symmetry \cite{Mars-Seno,Vera2}. It would be desirable to have more results along those lines, including, in particular, the matching of two non-vacuum solutions. 

As a step in this direction, we consider families of stationary spacetimes. For these spacetimes one can perform a $3+1$ splitting using the integral curves of the timelike Killing vector field. The Einstein equations can then be rewritten as equations determining two vector fields (one of which is a gradient) and a Riemannian metric on the quotient $3$-manifold, in what has been called the quasi-Maxwell formalism \cite{LN98, NZ97, O02}. While equivalent to the standard formulation, this approach allows a better intuitive understanding of some aspects of stationary spacetimes.

We will use the quasi-Maxwell formalism in order to prove general results for the matching of stationary spacetimes with additional symmetries. The matching will be performed along a stationary timelike hypersurface, so that the integral curves of the two timelike Killing vector fields coincide on the matching hypersurface. 

The organization of the paper is as follows. After briefly reviewing the quasi-Maxwell formalism (Section 1), we write the matching conditions using the gravitational and gravitomagnetic fields on the space manifold (Section 2). We then use the insight provided by this $3+1$ decomposition of the matching conditions to prove new results for the matching of stationary perfect fluid spacetimes with spherical, planar, hyperbolic and cylindrical symmetry (Section 3). We also recover particular examples (see \cite{Debbasch} and references therein) where the existence (but not the uniqueness) of matching solutions in stationary symmetry has been shown before. Finally, we use the quasi-Maxwell formalism to find an explicit interior for the cylindrical analogue of the NUT spacetime (Section 4). 

We use units such that $c=G=1$ and take Latin indices $i,j,\ldots$ to run from $1$ to $3$.
\section{Quasi-Maxwell formulation for stationary spacetimes}
In this section we briefly review the quasi-Maxwell formalism for stationary spacetimes. 
For more details see \cite{O02}.

Recall that a {\em stationary spacetime} $(M,g)$ is a Lorentzian $4$-manifold with a global timelike Killing vector field $T$. We assume that $T$ is complete and that the $\bbR$-action determined by its flow is free and proper\footnote{This will happen if for instance $(M,g)$ is chronological \cite{Harris}.}. The quotient space $\Sigma=M/\bbR$, which can be thought of as the space of all stationary observers, is then a $3$-dimensional manifold, called the {\em space manifold}. Moreover, the quotient map $\pi:M \to \Sigma$ is a submersion, and hence $M$ is a principal $\bbR$-bundle over $\Sigma$. Since all principal $\bbR$-bundles are trivial, we have $M \cong \bbR \times \Sigma$ \cite{Anderson, Beig-Schmidt}. The choice of global trivialization for $M$ is, of course, not unique, and amounts to choosing a map $t:M \to \bbR$ such that $T=\frac{\partial}{\partial t}$. Note that any two such maps will differ by a function $f:\Sigma \to \bbR$. If $\{x^i\}$ are local coordinates on $\Sigma$, we can write the line element of $(M,g)$ as
\begin{equation} \label{metric}
ds^2= -e^{2\phi}\left(dt+A_idx^i\right)^2+\gamma_{ij}dx^idx^j
\end{equation}
where the functions $\phi$, $A_i$ and $\gamma_{ij}$ do not depend on the coordinate $t$. Therefore the covariant tensor fields $\phi$, $A=A_idx^i$ and $\gamma=\gamma_{ij}dx^i\otimes dx^j$ are pull-backs by $\pi$ of the covariant tensor fields on $\Sigma$ with the same expressions, which we denote by the same symbols. Clearly $\gamma$ is a Riemannian metric in $\Sigma$, independent of the choice of $t$. It yields the radar distance measured between nearby stationary observers \cite{LL97}. The differential forms $G=-d \phi$ and $H=-e^ \phi dA$ are also independent of this choice (as $t \mapsto t+f$ gives $\phi \mapsto \phi$ and $A \mapsto A+df$). Moreover, $\gamma, G$ and $H$ are also invariant under the timelike Killing vector field rescaling $T \mapsto e^c T$, which gives $\phi \mapsto \phi + c$ and $A \mapsto e^{-c}A$. We define the {\em gravitational} and {\em gravitomagnetic} vector fields $\GG$ and $\HH$ through
\begin{align}
& G=\gamma(\GG,\cdot) \\
& H=\epsilon(\HH,\cdot,\cdot)
\end{align}
where $\epsilon$ is a Riemannian volume form on $(\Sigma, \gamma)$ (which we assume to be orientable).

Each vector ${\bf v} \in T_p\Sigma$ determines a unique vector field $\widetilde{\bf v}$ along the fiber $\pi^{-1}(p) \subset M$ satisfying $g(T,\widetilde{\bf v})=0$ and $\pi_* \widetilde{\bf v} = {\bf v}$. One can show that $\widetilde{\GG}$ is just minus the acceleration of the stationary observers, and that $\widetilde{\HH}$ is twice their vorticity. Moreover, the metric on the space manifold is given by $\gamma({\bf v}, {\bf w})=g(\widetilde{\bf v},\widetilde{\bf w})$ 
 
Let $\nabla$ be the Levi-Civita connection of $(\Sigma, \gamma)$ and let $\widetilde{\nabla}$ be the Levi-Civita connection of $(M,g)$. If $u$ represents the unit tangent vector to a timelike geodesic then the motion equation
\[
\widetilde{\nabla}_{u}\,u=0
\]
is equivalent to
\[
\nabla_{\uu}\,\uu= u^0\left( u^0\,\GG+\uu\times\HH \right)
\]
where ${\bf u}=\pi_* u$, $u^0 = \left( 1 + \uu^2 \right)^\frac12$ (with $\uu^2 = \gamma(\uu,\uu)$) and $\times$ is the cross product defined by $\gamma$ and $\epsilon$. There is an obvious analogy with the equation of motion of a charged particle on an electromagnetic field \cite{O02}, justifying the designations for $\GG$ and $\HH$.

This analogy can be pushed further by writing the Einstein equations for a perfect fluid with density $\mu$, pressure $p$ and $4$-velocity $u$: if $R_{ij}$ and $\nabla_iG_j$ represent the components of the Ricci tensor of $\nabla$ and of the covariant derivative of $G$, these reduce to the {\em quasi-Maxwell equations}
\begin{align*}
& \dive\GG= \GG^2+\frac12\HH^2-8\pi(\mu+p)\uu^2-4\pi(\mu+3p); \\
& \curl\HH=2\GG\times\HH-16\pi(\mu+p)u^0\uu;  \\
& R_{ij}+\nabla_iG_j=G_iG_j+\frac12 H_iH_j-\frac12\HH^2\gamma_{ij} + 8\pi\left((\mu+p)u_iu_j+\frac12(\mu-p)\gamma_{ij}\right). 
\end{align*}

These equations clearly show the role of $\mu + 3p$ as the source of the gravitational field when the fluid is at rest, as well as the nonlinear source terms for $\GG$ and $\HH$ akin to the field's energy density and Poynting vector\footnote{There are factors of $2$ with respect to the corresponding electromagnetic formulae, reflecting the spin $2$ nature of the gravitational interaction.}. Notice however that most of the equations are equations on the geometry of the space manifold.
\section{Matching in the quasi-Maxwell formalism}
Let $\sigma\subset \Sigma$ be a connected two-sided (hence orientable) embedded surface with unit normal vector field ${\bf n}$, which can be assumed to extend to a neighbourhood $U$ of $\sigma$ as a unit geodesic field. Then $\widetilde{\sigma} = \pi^{-1}(\sigma) \subset M$ is a timelike hypersurface with unit normal vector field $\widetilde{\bf n}$, which is a unit geodesic field on $\pi^{-1}(U)$. In local coordinates, if
\[
{\bf n}=n^i\frac{\partial}{\partial x^i}
\]
then
\[
\widetilde{\bf n} = n^i\frac{\partial}{\partial x^i}-g\left(T,n^i\frac{\partial}{\partial x^i}\right) \frac{T}{g(T,T)}= n^i\frac{\partial}{\partial x^i} - (A_i n^i)\frac{\partial}{\partial t}
\]
The second fundamental form of $\widetilde{\sigma}$ is
\begin{align}
\widetilde{K} & = \frac12 \pounds_{\widetilde{\bf n}} g = - d \phi \left(\widetilde{\bf n}\right) \left(\omega^0\right)^2 - e^{\phi} \omega^0 \left(d\left(-A_i n^i\right) + \iota\left(\widetilde{\bf n}\right) dA + d\left(A_i n^i\right) \right) + K \nonumber \\
& = \gamma(\GG,{\bf n}) \left(\omega^0\right)^2 + (\iota({\bf n}) H)\, \omega^0 + K, \label{extrinsic} 
\end{align}
where
\[
\omega^0 = e^{\phi} \left(dt+A_idx^i\right)
\]
and
\[
K = \frac12 \pounds_{\bf n} \gamma
\]
is the second fundamental form of $\sigma$. In this calculation we have used the Cartan formula
\[
\pounds_X \theta = d(\iota(X)\theta) + \iota(X) d\theta,
\]
valid for any vector field $X$ and differential form $\theta$, and identified covariant tensors on $\Sigma$ with their pull-backs by $\pi:M \to \Sigma$.

If we are to match $(M,g)$ to another stationary spacetime across the stationary hypersurface $\widetilde{\sigma}$ we must match $g_{|_{\widetilde{\sigma}}}$ and $\widetilde{K}$. If in addition we do the matching so that the integral lines of the timelike Killing vector fields coincide, it is clear from \eqref{metric} that matching $g_{|_{\widetilde{\sigma}}}$ is equivalent to matching $\phi_{|_\sigma}$ (up to a constant, corresponding to the freedom associated to the rescaling $T \mapsto e^c T$), $A_{|_\sigma}$ (up to $df_{|_\sigma}$ for some function $f:\Sigma \to \bbR$, corresponding to the freedom associated to the choice of $t$) and $\gamma_{|_\sigma}$. Since $\sigma$ is connected, matching $\phi_{|_\sigma}$ up to a constant is equivalent to matching $d\phi_{|_\sigma}$, which in turn is equivalent to matching the tangential component of $\GG$. On the other hand, matching $A_{|_\sigma}$ up to $df_{|_\sigma}$ is equivalent to matching $dA_{|_\sigma}$\footnote{At least locally; if $\sigma$ is not simply connected then we must also match the integrals $\oint_\Gamma A$ of $A$ along the generators $\Gamma$ of the fundamental group $\pi_1(\sigma)$.}. Given that $\phi_{|_\sigma}$ is matched, this is the same as matching $H_{|_\sigma}$. If ${\bf v}, {\bf w}$ are tangent to $\sigma$ then
\[
H_{|_\sigma}({\bf v}, {\bf w}) = \epsilon(\HH,{\bf v}, {\bf w}) = \gamma(\HH,{\bf v} \times {\bf w}). 
\]
Since ${\bf v} \times {\bf w}$ is orthogonal to $\sigma$, we see that matching $H_{|_\sigma}$ is equivalent to matching the normal component of $\HH$. Similarly, it is clear from \eqref{extrinsic} that matching $\widetilde{K}$ is the same thing as matching  $\gamma(\GG,{\bf n})$, $(\iota({\bf n}) H)_{|_\sigma}$ and $K$. Now $\gamma(\GG,{\bf n})$ is simply the normal component of $\GG$. On the other hand, if ${\bf v}$ is tangent to $\sigma$ then
\[
(\iota({\bf n}) H)_{|_\sigma}({\bf v}) = H({\bf n}, {\bf v}) = \epsilon(\HH,{\bf n}, {\bf v}) = \gamma(\HH,{\bf n} \times {\bf v}).
\]
Since any vector ${\bf w}$ tangent to $\sigma$ can be obtained in the form ${\bf w} = {\bf n} \times {\bf v}$ for an appropriate tangent vector ${\bf v}$, we see that matching $(\iota({\bf n}) H)_{|_\sigma}$ is the same thing as matching the tangential component of $\HH$. 

Thus we have proved the following result.

\begin{Teo} \label{main}
Let $(M^-,g^-)$ and $(M^+,g^+)$ be stationary spacetimes with complete timelike Killing vector fields inducing free and proper $\bbR$-actions, and assume that the corresponding space manifolds $\Sigma^-$ and $\Sigma^+$ are orientable. Let $\sigma^- \subset \Sigma^-$ and $\sigma^+ \subset \Sigma^-$ be diffeomorphic topologically closed two-sided surfaces with unit normal vector fields ${\bf n}^-$ and ${\bf n}^+$, and let $f:\sigma^- \to \sigma^+$ be a diffeomorphism. Then $(M^-,g^-)$ and $(M^+,g^+)$ can be matched by a lift $\widetilde{f}:\widetilde{\sigma}^- \to \widetilde{\sigma}^+$ of $f$ if and only if:
\begin{enumerate}
\item [(i)]$f$ carries the induced metric on $\sigma^-$ to the induced metric on $\sigma^+$;
\item [(ii)] $f$ carries the second fundamental form of $\sigma^-$ to the second fundamental form of $\sigma^+$;
\item [(iii)] $f$ carries the gravitational and gravitomagnetic fields ${\bf G}^-$ and ${\bf H}^-$ on $\sigma^-$ to the corresponding fields ${\bf G}^+$ and ${\bf H}^+$ on $\sigma^+$ under the identification $T\Sigma^-|_{\sigma^-} \cong T\Sigma^+|_{\sigma^+}$ determined by $f$, ${\bf n}^-$ and ${\bf n}^+$.
\end{enumerate}
\end{Teo}

Informally, stationary spacetimes can be matched along stationary hypersurfaces if and only if the corresponding surfaces on the space manifold have the same metric, second fundamental form and gravitational and gravitomagnetic fields. Note the similarity with electromagnetostatics, where the junction conditions are the continuity of the electric and magnetic fields.\footnote{There are also similarities with the $2+1$ decomposition of the matching conditions in \cite{Mars}.}
\section{Local existence and uniqueness of stationary matchings}
In this section we study the problem of existence and uniqueness of families of matchings of stationary spacetimes with additional symmetries. In all cases the matching is performed along hypersurfaces which preserve the symmetries (see \cite{Vera} for general geometrical and algebraic considerations about symmetry-preserving matchings). 
\subsection{Spherical, Toroidal and Higher Genus  $\Lambda$-Stars}
We shall start with the static cases. 
The line element for a static perfect fluid with spherical ($k=1$), planar ($k=0$) or hyperbolic ($k=-1$) symmetry can be written as \cite{Schutz, Kramer}
\[
ds^2 = - e^{2 \phi(r)} dt^2 + \frac{dr^2}{k-\frac{2m(r)}r} + r^2 (d \theta^2 + \Sigma^2(k,\theta) d \varphi^2)
\]
where
\[
\Sigma(k,\theta) = 
\begin{cases} 
\sin \theta  &\text{ if } k=1 \\
\theta  &\text{ if } k=0 \\
\sinh \theta &\text{ if } k=-1 \\
\end{cases}
\]
and the functions $\phi(r)$ and $m(r)$ are related to the rest energy density $\mu(r)$ and the rest pressure $p(r)$ by
\begin{align}
& \frac{dm}{dr} = 4 \pi r^2 \mu; \label{mass} \\
& \frac{dp}{dr} = - \frac{(\mu + p)(m + 4\pi r^3 p)}{r(kr-2m)}; \label{pressure} \\
& \frac{d\phi}{dr} = \frac{m + 4\pi r^3 p}{r(kr-2m)}. \label{potential}
\end{align}
Notice that we must have
\[
k-\frac{2m(r)}r > 0,
\]
and hence $m(r)<0$ for $k=0$ or $k=-1$. This can be accommodated by a negative cosmological constant $\Lambda$, in terms of which the total rest energy density and rest pressure can be written
\begin{align*}
& \mu = \mu_{matter} + \frac{\Lambda}{8\pi}; \\
& p = p_{matter} - \frac{\Lambda}{8\pi}.
\end{align*}
The space metric for these spacetimes is
\[
\gamma = \frac{dr^2}{k-\frac{2m(r)}r} + r^2 (d \theta^2 + \Sigma^2(k,\theta) d \varphi^2)
\]
and the gravitational field is
\[
{\bf G} = - \phi'(r) \left( k-\frac{2m(r)}r \right) \frac{\partial}{\partial r}.
\]
We can match two of these spacetimes across a surface of constant $r$. The outward normal to the corresponding surface on the space manifold is
\[
{\bf n} = \left( k-\frac{2m(r)}r \right)^\frac12 \frac{\partial}{\partial r},
\]
yielding the second fundamental form
\[
K = \frac12 \pounds_{\bf n} \gamma = r \left( k-\frac{2m(r)}r \right)^\frac12 (d \theta^2 + \sin^2 \theta d \varphi^2).
\]
Matching of the induced metric on the constant $r$ surfaces implies continuity of the $r$ coordinate function. Matching of the second fundamental form then yields the continuity of the mass function $m(r)$. Finally, matching of the gravitational fields requires that $\phi'(r)$, and hence the pressure $p(r)$, should also match.
\begin{Prop}
Two spacetimes with a static perfect fluid and spherical\footnote{See \cite{Fayos-Sen-Torres} for the general matching in spherical symmetry.}, planar or hyperbolic symmetry can be matched across a surface of symmetry if and only if the radius, mass and pressure functions are continuous at the surface. 
\end{Prop}
We can construct models of spherical, toroidal and higher genus stars by matching two of these spacetimes (after taking the appropriate quotients by discrete subgroups of the plane or hyperbolic symmetry groups). The exterior will be the Kottler solution, corresponding to the case when the perfect fluid is simply vacuum plus a cosmological constant. It is given by the mass function
\[
m(r) = m_0 + \frac{\Lambda}6 r^3
\]
(where $m_0$ is a constant), and satisfies
\[
e^{2 \phi(r)} = k-\frac{2m(r)}r.
\]
Notice that the Kottler solution has two free parameters, $m_0$ and $\Lambda$, which can be used to match $m(r)$ and $p(r)$ uniquely\footnote{It is easily checked that for fixed $r$ the map $(m_0,\Lambda) \mapsto (m_0 + \frac{\Lambda}6 r^3,-\frac{\Lambda}{8\pi})$ is a bijection.}. Therefore we have the following result.
\begin{Prop}
A spacetime with a static perfect fluid and spherical, planar or hyperbolic symmetry can be matched across a surface of symmetry to a unique Kottler solution.
\end{Prop}
The star's interior can be obtained by picking an equation of state and integrating \eqref{mass}, \eqref{pressure} and \eqref{potential}. For instance, the equation of state $\mu = \text{constant}$ yields
\begin{align}
& m = \frac{4 \pi}3 r^3 \mu; \\
& \left| \frac{\mu + 3p}{\mu + p} \right| = c \left| k - \frac{2m}{r} \right|^\frac12, \label{pressure2}
\end{align}
where $c$ is an integration constant. Interestingly, toroidal ($k=0$) stars with constant density do not exist for physically reasonable matter ($\mu_{matter}, p_{matter} > 0$), as \eqref{pressure2} requires that at the center $r=0$ one should have
\[
\mu + 3p = 0 \Leftrightarrow \mu_{matter} + 3 p_{matter} = \frac{\Lambda}{4\pi} < 0.
\]
Moreover, \eqref{pressure} shows that for toroidal stars $p$ always diverges at $r=0$ unless $\mu+p$ vanishes at this point. There are similar difficulties in constructing higher genus stars \cite{SM98}.

For static solutions, the quasi-Maxwell formulation of the matching conditions is not particularly advantageous. However, it still can provide valuable insights. For instance, consider the problem of matching the Einstein universe, obtained by taking $k=1$, $\phi=0$ and
\[
m(r) = \frac{\Lambda}2 r^3,
\]
to the Kottler solution. Since we have $\GG = {\bf 0}$ throughout the Einstein universe, we must have $\GG={\bf 0}$ at the matching surface; this uniquely picks a surface on the Kottler solution, given by
\[
\frac{d}{dr} e^{2 \phi(r)} = 0 \Leftrightarrow r^3 = \frac{3m_0}{\Lambda}.
\]
\subsection{Rigidly rotating perfect fluids}
The line element for a stationary, cylindrically symmetric spacetime containing a rigidly rotating perfect fluid  can be cast in the form \cite{Kramer}
\begin{equation}
ds^2=-\frac{1}{h^2}(dt+\rho d\varphi)^2+\frac{1}{EBh}d\rho^2+\frac{E}{h^2}d\varphi^2+\frac{h^3}{B}dz^2, \label{metricfluid}
\end{equation}
where
\begin{equation}
B(\rho)=h^5(\rho) E^{-1}(\rho)\exp\left(\int \rho E^{-1}(\rho)d\rho\right) \label{Bfluid}
\end{equation}
for $E=E(\rho)$ and $f=h^3(\rho)$ satisfying
\begin{equation}
E^2 f''+(\rho E-EE')f'-\frac{3}{4}(EE''-E'^2+\rho E'-E)f=0, \label{eqfluid}
\end{equation}
where a prime denotes differentiation with respect to $\rho$.
The fluid is moving along the integral curves of $\partial / \partial t$, and its pressure and mass density are given by
\begin{align}
& 8\pi p=Bh^{-5}E^{-1}\left[\frac{1}{4}f^2(E-\rho E'+E'^2)+\frac{1}{3}Ef'(\rho f+Ef'-2E'f)\right];\label{pressurefluid}\\
& 8\pi(\mu+p)= Bh(1-E''/2).\label{densityfluid}
\end{align}
In this case, we obtain
\begin{align}
& {\bf G} = EBh' \frac{\partial}{\partial \rho}; \nonumber \\
& {\bf H} = - \frac{B}{h}\frac{\partial}{\partial z}; \label{Hfluid} \\
& \gamma = \frac{1}{EBh}d\rho^2+\frac{E}{h^2}d\varphi^2+\frac{h^3}{B}dz^2. \label{gammafluid}
\end{align}
The normal vector to a cylindrical surface $\sigma = \{\rho = \text{constant} \}$ is
\[
{\bf n} = \sqrt{EBh} \frac{\partial}{\partial \rho},
\]
and the extrinsic curvature on $\sigma$ is then
\begin{equation}
K= \frac{1}{2}\sqrt{EBh}\left(\frac{E}{h^2}\right)'d\varphi^2+\frac{1}{2}\sqrt{EBh}\left(\frac{h^3}{B}\right)'dz^2. \label{Kfluid}
\end{equation}
Notice also that
\begin{equation}
\GG = h' \sqrt{\frac{EB}{h}} {\bf n}. \label{Gfluid}
\end{equation}

\begin{Prop}
Two stationary, cylindrically symmetric spacetimes (\ref{metricfluid}) containing rigidly rotating perfect fluids can be matched along a stationary cylindrical surface $\{\rho = \text{constant} \}$ if and only if the functions $\rho,E,E',f,f'$ and $B$ are continuous on this surface.
\end{Prop}
{\bf Proof:} From \eqref{Hfluid} and \eqref{gammafluid} we see that matching $\HH$ and $\gamma_{|_\sigma}$ is equivalent to matching $h,E$ and $B$. From \eqref{Gfluid} we see that matching $\GG$ is the same as matching $h'$. Finally, from \eqref{Kfluid} we see that matching $K$ is the same as matching $E'$ and $B'$. Since $f=h^3$, matching $h$ and $h'$ is equivalent to matching $f$ and $f'$. On the other hand, from \eqref{Bfluid} we see that
\[
\frac{B'}{B} = \frac{5h'}{h} - \frac{E'}{E} + \frac{\rho}{E},
\]
and thus once one has matched $E'E',f,f'$ and $B$, matching $B'$ is the same as matching $\rho$. $\square$

\begin{Rem}
Notice from \eqref{pressurefluid} and \eqref{densityfluid} that the pressure (but not the density) must also be continuous. This is an instance of the Israel matching conditions for the energy-momentum tensor.
\end{Rem}
The line element \eqref{metricfluid} contains vacuum and dust solutions as particular cases. More precisely, we have the following result.

\begin{Prop}
The line element \eqref{metricfluid} represents a $\Lambda$-vacuum solution if and only if 
\[
E(\rho) = \rho^2 + \alpha\rho + \beta
\] for some constants $\alpha, \beta \in \bbR$. The same line element represents a $\Lambda$-dust solution if and only if $f$ is constant and 
\[
E(\rho) = \alpha e^{\beta \rho} - \frac{\beta \rho + 1}{\beta^2}
\]
for some constants $\alpha, \beta \in \bbR$ (with $\beta \neq 0$).
\end{Prop}
{\bf Proof:} If \eqref{metricfluid} represents a $\Lambda$-vacuum solution then $\mu + p = 0$, which by \eqref{densityfluid} is equivalent to $E''(\rho)=2$. On the other hand, if $E''(\rho) = 2$ then the formula \cite{Kramer}
\[
8 \pi p = \int Bh'(1-E''/2) d\rho
\]
implies that $p$ is constant. 

If \eqref{metricfluid} represents a $\Lambda$-dust solution then we must have $\GG={\bf 0}$ (because in this case the stationary observers, which are comoving with the fluid, are not accelerating), and hence $h$ (or $f$) must be constant. Equation \eqref{eqfluid} then yields
\[
EE''-E'^2+\rho E'-E = 0,
\]
which is readily solved to give the expression above. On the other hand, substituting this expression in \eqref{Bfluid}, it is easy to show from \eqref{pressurefluid} (under the assumption that $f$ is constant) that $p$ is constant. $\square$

\begin{Teo}
A stationary, cylindrically symmetric spacetime containing a rigidly rotating perfect fluid can be matched along a stationary cylindrical surface $\{\rho = c \}$ to a $\Lambda$-vacuum solution, or, if $f'(c)=0$, to a $\Lambda$-dust solution. These matchings are unique within this class of metrics.
\end{Teo}
{\bf Proof:}
Fixing the surface $\{\rho = c \}$ on a given rigidly rotating perfect fluid determines the values of $\rho,E,E',f,f'$ and $B$. Since we have two free parameters on the expression of $E(\rho)$ for $\Lambda$-vacuum or $\Lambda$-dust solutions, we can use these to match the values of $E$ and $E'$ \footnote{It is easily checked that for fixed $\rho$ the map $(\alpha,\beta) \mapsto (E(\rho),E'(\rho))$ is a bijection.}. It is also easy to match the value of $B$, which from \eqref{Bfluid} is seen to be defined up to a multiplicative constant. Finally, in the $\Lambda$-vacuum case the values of $f$ and $f'$ provide initial data for \eqref{eqfluid}, which can be seen as a second order linear ODE for $f$. Solving this ODE (which admits a global solution) yields a unique local solution (defined on $\{ f(\rho) > 0 \}$). In the $\Lambda$-dust case the matching of $f$ and $f'$ is trivially unique. $\square$

\begin{Rem}
The above theorem includes, for instance, the matching of the van Stockum dust to the Lewis vacuum \cite{vanStockum, Tipler, Bonnor1}, or the matching of the G\"odel $\Lambda$-dust to the Lewis $\Lambda$-vacuum generalization \cite{Bonnor2}, which in particular are unique within this class of metrics.
\end{Rem}

\section{NUT matchings}
In this section, using the quasi-Maxwell formalism, we revisit a matching of a perfect fluid with the NUT spacetime and then derive a new solution which results from the matching of a line monopole-like dust with the cylindrical analogue of the NUT spacetime.
\subsection{Matching a perfect fluid with NUT}
Recall that the NUT spacetime is a stationary solution of the vacuum Einstein field equations describing a gravitomagnetic monopole, given in local coordinates by\footnote{Here we choose to regard the NUT spacetime as $\bbR \times \Sigma$, where $\Sigma$ is diffeomorphic to $\bbR^3$ minus a singular line, so that Theorem \ref{main} applies.}
\[
{ds^2}^+ = - e^{2\phi} (dt + 2l \cos \theta d\varphi)^2  + e^{-2\phi} dr^2 + (r^2 + l^2) (d \theta^2 + \sin^2 \theta d \varphi^2),
\]
where
\[
e^{2\phi} = 1 - 2 \frac{Mr + l^2}{r^2 + l^2}
\]
and $M,l$ are two parameters representing the total mass and (half) the gravitomagnetic charge \cite{NUT63, LN98, Kramer}. The metric of the space manifold determined by the stationary observers is
\[
\gamma^+ = e^{-2\phi} dr^2 + (r^2 + l^2) (d \theta^2 + \sin^2 \theta d \varphi^2),
\]
and the gravitomagnetic potential $1$-form is
\[
A^+ = 2l \cos \theta d\varphi.
\]
Thus the gravitomagnetic field corresponds to the $2$-form
\[
H^+ = - e^\phi dA^+ = 2l e^\phi \sin \theta d\theta \wedge d \varphi,
\]
and is therefore the radial vector field
\[
{\bf H}^+ = \frac{2le^{2\phi}}{r^2 + l^2} \frac{\partial}{\partial r}.
\]
Moreover, the gravitational field corresponds to the $1$-form
\[
G^+ = - \phi' dr,
\]
and so
\[
{\bf G}^+ = - \phi' e^{2\phi} \frac{\partial}{\partial r}.
\]
A particularly simple interior for the NUT spacetime was given in \cite{Gergely} as
\[
{ds^2}^- = - \sin^4 \psi \left(dt + 2R \cos \theta d \varphi - \frac{R}{\sin^2 \psi} d\psi \right)^2 + R^2 (d \psi^2 + \sin^2 \psi (d \theta^2 + \sin^2 \theta d \varphi^2)).
\]
This stationary solution describes a perfect fluid with constant density
\[
\mu = \frac6{R^2}
\]
and pressure
\[
p = \frac4{R^2 \sin^2 \psi} - \frac6{R^2}.
\]
Notice that the space manifold is a round $S^3$ of radius $R$. The gravitational field corresponds to the $1$-form
\[
G^- = - d \log \sin^2 \psi = - \frac{2 \cos \psi}{\sin \psi} d \psi,
\]
and therefore is 
\[
\GG^- = - \frac{2 \cos \psi}{R^2 \sin \psi} \frac{\partial}{\partial \psi}.
\]
The gravitomagnetic field corresponds to the $2$-form
\[
H^- = - \sin^2 \psi d (2R \cos \theta d \varphi) = 2 R \sin^2 \psi \sin \theta d\theta \wedge d \varphi,
\]
and consequently is
\[
\HH^- = \frac{2}{R^2} \frac{\partial}{\partial \psi}.
\]
Notice that the solution has singularities at the ``poles'' $\sin \psi = 0$, where the pressure diverges. The gravitational field also diverges at the poles, and vanishes at the ``equator'' $\psi = \frac{\pi}2$. The gravitomagnetic field remains bounded (and in fact the fluid has constant vorticity $\frac1R$). Also, both fields are aligned along the integral lines of $\frac{\partial}{\partial \psi}$, which makes it plausible that this spacetime can be matched along a surface $\psi = \text{constant}$ to the NUT spacetime along a surface $r = \text{constant}$. For these surfaces, we have
\[
K^- = \left( \pounds_{\frac1R \frac{\partial}{\partial \psi}} \gamma^- \right)_{|_\sigma} = \frac{\cos \psi}{R \sin \psi} R^2 \sin^2 \psi (d \theta^2 + \sin^2 \theta d \varphi^2)
\]
and
\[
K^+ = \left( \pounds_{e^\phi \frac{\partial}{\partial r}} \gamma^+ \right)_{|_\sigma} = \frac{re^\phi}{r^2 + l^2} (r^2 + l^2) (d \theta^2 + \sin^2 \theta d \varphi^2).
\]
The matching conditions can be immediately written as
\[
\begin{cases}
\GG^- \eqq \GG^+ \\
\HH^- \eqq \HH^+ \\
({\gamma^-})_{|_\sigma} \eqq (\gamma^+)_{|_\sigma} \\
K^- \eqq K^+
\end{cases}
\Leftrightarrow
\begin{cases}
- \frac{2 \cos \psi}{R \sin \psi} \eqq - \phi' e^\phi \\
\frac{2}{R} \eqq \frac{2le^{\phi}}{r^2 + l^2}  \\
R^2 \sin^2 \psi \eqq r^2 + l^2 \\
\frac{\cos \psi}{R \sin \psi} \eqq \frac{re^\phi}{r^2 + l^2}
\end{cases}
\Leftrightarrow
\begin{cases}
r \eqq l \cot \psi \\
l \eqq R \sin^2 \psi \\
M \eqq -l \tan \psi \\
\cot \psi \eqq \frac{\sqrt{2}}{2}
\end{cases}
\]
(where $\eqq$ means equality on $\sigma$), and are easily seen to be equivalent to the ones in \cite{Gergely}.
\subsection{Matching a Line Monopole-like dust with Cylindrical NUT}
We now consider the problem of finding an interior for the cylindrical analogue of the NUT spacetime, given by the line element \cite{NZ97}
\[
{ds^2}^+=-e^{2\phi}(dt-Lzd\varphi)^2+e^{-2\phi}\left(\frac{2ma}{L}\rho^{2m^2}(d\rho^2+dz^2)+\rho^2d\varphi^2\right)
\]
where
\[
e^{2\phi} = \frac{2m}{L}\frac{1}{\cosh{(2m \ln (\rho/c))}}
\]
and $m, a, L$ and $c$ are constants. This metric is a generalization of the Levi-Civita cylindrically symmetric static metric in the same way that the NUT metric is a vacuum generalization of Schwarzschild. In this case, we get
\begin{align*}
& \gamma^+=e^{-2\phi}\left(\frac{2ma}{L}\rho^{2m^2}(d\rho^2+dz^2)+\rho^2d\varphi^2\right); \\
& G^+ = \frac{m}{\rho}\tanh {(2m\ln(\rho/c))}d\rho; \\
& \HH^+ = -\frac{L^2e^{4\phi}}{2ma\rho^{2m^2+1}} \frac{\partial}{\partial \rho}.
\end{align*}
A possible interior can be obtained by applying the Ehlers transformation \cite{Ehlers, Kramer} to the Levi-Civita solution, given by the line element \cite{Kramer, Wang}
\[
ds^2=-r^{2 \alpha} dt^2+r^{-2 \alpha}\left(r^{2 \alpha^2}(dr^2 + dz^2) + k^2 r^2 d \varphi^2 \right).
\]
In terms of the quasi-Maxwell formalism, the Ehlers transformation associates to each static vacuum solution of the Einstein field equations, described by the gravitational potential $\psi$ and the space metric $\hat{\gamma}$, a stationary solution with
\[
\gamma = e^{2 \psi} \hat{\gamma}, \quad \GG = {\bf 0} \quad \text{ and } \quad \HH = 2\grad \psi,
\]
whose matter content is dust with density\footnote{This can be checked directly from the quasi-Maxwell equations.}
\[
\mu = \frac1{8 \pi} \HH^2.
\]
Therefore, we will take as the interior the stationary solution characterized by
\begin{align*}
& \gamma^-=r^{2\alpha^2}(dr^2+dz^2) + k^2 r^2d\varphi^2; \\
& \GG^- = {\bf 0}; \\
& \HH^- = \frac{2\alpha}{r^{2 \alpha^2 + 1}} \frac{\partial}{\partial r},
\end{align*}
corresponding to a freely falling dust with density
\[
\mu = \frac{\alpha^2}{2 \pi r^{2 \alpha^2 + 2}}.
\]
Notice that for $\alpha \neq 0$ there is a singularity at $r=0$, where both the mass density and the vorticity blow up.

The condition 
\[
\GG^+ \eqq \GG^- = {\bf 0}
\]
immediately singles out the cylindrical surface $\sigma = \{\rho = c \}$ as the matching surface on the exterior. Accordingly, we choose $\sigma$ to be a surface of constant $r$ in the interior. The normal vectors to these surfaces are
\begin{align*}
& {\bf n}^- = r^{-\alpha^2} \frac{\partial}{\partial r};\\
& {\bf n}^+ = \frac1{\sqrt{a}} \rho^{-m^2} \frac{\partial}{\partial \rho},
\end{align*}
and the induced metrics are
\begin{align*}
& (\gamma^-)_{|_\sigma} = k^2 r^2 d\varphi^2 + r^{2\alpha^2} dz^2; \\
&(\gamma^+)_{|_\sigma} = \frac{L}{2m} \rho^2 d\varphi^2 + a\rho^{2m^2} dz^2.
\end{align*}
The condition $(\gamma^-)_{|_\sigma} \eqq (\gamma^+)_{|_\sigma}$ then yields
\begin{equation}
k^2 r^2 \eqq \frac{L}{2m} \rho^2 \quad \text{ and } \quad r^{2\alpha^2} \eqq a\rho^{2m^2}. \label{cylNUTmet}
\end{equation}
Moreover, since $\frac{\partial}{\partial \varphi}$ and $\frac{\partial}{\partial z}$ are Killing vector fields, we have
\begin{align*}
& K^-_{\varphi\varphi} = \frac12 \nabla_{{\bf n}^-} \gamma^-_{\varphi \varphi} = k^2 r^{1-\alpha^2} ; \\ 
& K^-_{zz} = \frac12 \nabla_{{\bf n}^-} \gamma^-_{zz} = \alpha^2 r^{\alpha^2-1} ; \\ 
&K^+_{\varphi\varphi} = \frac12 \nabla_{{\bf n}^+} \gamma^+_{\varphi \varphi} = \frac{L}{2m\sqrt{a}} \rho^{1-m^2} ; \\
& K^+_{zz} = \frac12 \nabla_{{\bf n}^+} \gamma^+_{zz} =  \sqrt{a} m^2 \rho^{m^2-1}. 
\end{align*}
The condition $K^- \eqq K^+$ then yields
\begin{equation}
k^2 r^{1-\alpha^2} \eqq \frac{L}{2m\sqrt{a}} \rho^{1-m^2} \quad \text{ and } \quad \alpha^2 r^{\alpha^2-1} \eqq \sqrt{a} m^2 \rho^{m^2-1}. \label{cylNUText}
\end{equation}
From \eqref{cylNUTmet} and \eqref{cylNUText} one readily obtains
\begin{align} \label{cylNUTconds}
r \eqq \rho, \quad \alpha^2 = m^2, \quad k^2 = \frac{L}{2m} \quad \text{ and } \quad a = 1.
\end{align}
Finally, the condition $\HH^- \eqq \HH^+$ yields
\[
\alpha \eqq - \frac{L^2e^{4\phi}}{2m}.
\]
Since 
\[
\rho \eqq c \Rightarrow e^{2 \phi} \eqq \frac{2m}{L},
\]
we obtain
\[
\alpha = -m,
\]
which implies the second condition in \eqref{cylNUTconds}. Thus we obtain a three-parameter family of matchings, parameterized by, say, $m, L$ and $c$. It is easily seen that there is a $1$-to-$1$ correspondence between interiors with a cylindrical surface singled out and exteriors with $a=1$.

One could think of trying the same approach on the spherically symmetric case, i.e.~matching the Ehlers transform of the Schwarzschild solution to the NUT solution. However this cannot possibly work, because $\GG$ never vanishes outside the horizon of the NUT spacetime. This is another example of the usefulness of the quasi-Maxwell formalism.
\section*{Acknowledgements}
FM thanks Marc Mars and Raul Vera for comments and references, Dep. Matem\'atica, Instituto Superior T\'ecnico for hospitality, FCT (Portugal) for grant SFRH/BPD/12137/2003 and CMAT, Universidade do Minho, for support. JN was partially supported by FCT (Portugal) through the Program POCI 2010/FEDER and by the grant POCI/MAT/58549/2004. 

\end{document}